\documentclass[showpacs,twocolumn,preprintnumbers,showkeys,superscriptaddress,amsmath,amssymb,nofootinbib]{revtex4-1}

\usepackage{color}
\usepackage{amsmath}
\usepackage{amssymb}
\usepackage[utf8]{inputenc}
\usepackage{amsfonts}
\usepackage{bm}
\usepackage{bbold}
\usepackage{graphics}
\usepackage{graphicx}

\newcommand{\be}{\begin{equation}}
\newcommand{\ee}{\end{equation}}
\newcommand{\ba}{\begin{eqnarray}}
\newcommand{\ea}{\end{eqnarray}}

\def\ni{\noindent}

\begin{document}

%

\title{\Large Vacuum material properties and Cherenkov radiation
  in Logarithmic Electrodynamics}

\author{Patricio Gaete} \email{patricio.gaete@usm.cl}
\affiliation{Departamento de F\'{i}sica and Centro Cient\'{i}fico-Tecnol\'ogico de Valpara\'{i}so-CCTVal,
Universidad T\'{e}cnica Federico Santa Mar\'{i}a, Valpara\'{i}so, Chile}

\author{J. A. Helay\"el-Neto}\email{helayel@cbpf.br}
\affiliation{Centro Brasileiro de Pesquisas F\'isicas, Rua Dr. Xavier Sigaud
150, Urca, Rio de Janeiro, Brasil, CEP 22290-180}

\date{\today}

\begin{abstract}
\ni
We study some observational signatures of nonlinearities of the electromagnetic field. First to all we show the vital role played by nonlinearities in triggering a material behavior of the vacuum with  $(\varepsilon > 0, \mu <0)$, which corresponds to a ferrimagnetic material. Secondly, the permittivity and susceptibility induced by nonlinearities are considered in order to obtain the refractive index via the dispersion relation
for logarithmic electrodynamics.
Finally, we consider the electromagnetic radiation produced by a moving charged particle interacting with  a medium characterized by nonlinearities of the electromagnetic field.
To this end we consider logarithmic electrodynamics. The result shows that the radiation is driven by the medium through which the particle travels like the one that happens in the Cherenkov effect.

\end{abstract}

\maketitle

\pagestyle{myheadings}
\markright{Vacuum material properties and Cherenkov radiation
  in Logarithmic Electrodynamics}

\section{Introduction}

The physical manifestations of vacuum electromagnetic nonlinearities have been a fascinating topic of research since the discovery by Euler and Heisenberg \cite{EH} of a striking prediction of quantum electrodynamics (QED), that is, the light-by-light scattering arising from the interaction of photons with virtual electron-positron pairs. As is well known, the physical consequences of this crucial finding, such as vacuum birefringence and vacuum dichroism, have been largely considered from different points of view \cite{Adler, Costantini,Ruffini,Dunne}. However, despite remarkable progress \cite{Bamber,Burke,Pike,Tommasini1,Tommasini2,PVLAS}, this prediction has not yet been confirmed. 

It is appropriate to remark, in this context, that recently the ATLAS and CMS collaborations at the Large Hadron Collider (LHC) have reported on the high energy gamma-gamma pair emission from virtual gamma-gamma scattering in ultraperipheral Pb-Pb collisions \cite{Atlas,Cms}. However, as emphasized in \cite{Robertson2}, in these results there is no modification of the optical properties of the vacuum. In addition, the coming of laser facilities has given rise  to various proposals to probe quantum vacuum nonlinearities \cite{Battesti, Ataman}. 
An interesting example is provided by the experiment (DeLLight project) \cite{Robertson2}, which exploits the change in the index of refraction due to nonlinear electrodynamics. 

In this connection, it may be recalled that different nonlinear electrodynamics of the vacuum may have significant contributions to photon-photon scattering such as Born-Infeld \cite{Born} and Lee-Wick \cite{Lee-Wick} theories. As is well known, these electrodynamics were introduced in order to avoid the divergences inherent in the Maxwell theory at short distances.

With these ideas in mind, in previous works \cite{Nonlinear,Logarithmic,Nonlinear2,Nonlinear3}, we have considered the physical effects presented by different models of $(3+1)$-D nonlinear electrodynamics in vacuum. Evidently, this has helped us to gain insights into the peculiarities of quantum vacuum nonlinearities in different contexts. For example, the Generalized Born-Infeld, and Logarithmic Electrodynamics the field energy of a point-like charge is finite, which also exhibit the vacuum birefringence phenomenon. As well as we have studied the lowest-order modifications of the static potential within the framework of the gauge-invariant but path-dependent variables formalism, which is an alternative to the Wilson loop approach. 

We further note that recently an interesting study on vacuum Cherenkov radiation in Euler-Heisenberg-like nonlinear electrodynamics has been considered in \cite{Macleod}. As is well known, a charged particle moving in a medium under an external electromagnetic field emits Cherenkov radiation when its velocity of light exceeds the phase velocity in that medium.

It is worth  recalling, at this stage, that any variation of the velocity of light with respect to $c = {1 \mathord{\left/
 {\vphantom {1 {\sqrt {{\varepsilon _0}{\mu _0}} }}} \right.
 \kern-\nulldelimiterspace} {\sqrt {{\varepsilon _0}{\mu _0}} }}$, where ${\varepsilon _0}$ and ${\mu _0}$ are the vacuum permittivity and the vacuum permeability respectively, is due to that light propagates in a medium. In this manner, we have to introduce the constants $\varepsilon$ and $\mu$ which characterize the medium. As is well known, the velocity of light in a medium is less than the velocity of light in vacuum by a factor (index of refraction) $n = {{\sqrt {\varepsilon \mu } } \mathord{\left/
 {\vphantom {{\sqrt {\varepsilon  \mu } } {\sqrt {{\varepsilon _0}{\mu _0}} }}} \right.
 \kern-\nulldelimiterspace} {\sqrt {{\varepsilon _0}\,{\mu _0}} }}$. 
From the previous remark it follows that both $\varepsilon$ and $\mu$ are positives. Nevertheless, as was first hypothesized in \cite{Veselago}, considerable attention has been paid recently to the $\varepsilon<0$ and $\mu<0$ case. The interest in studying this case is mainly due to the laboratory construction of an exotic form of dielectric, 
{\it metamaterial}, with both $\varepsilon$ and $\mu$ negatives. As it offers a valuably observational window on the constitutive parameters $\varepsilon$ and $\mu$, electrodynamics with 
metamaterial features has stimulated lots of experimental works \cite{Caloz}.

In this context it is particularly important to notice that, as emphasized in \cite{Caloz}, the four possible sign combinations in the pair $(\varepsilon, \mu)$ are $(+,+)$, $(+,-)$, $(+,-)$ and $(-,-)$. Evidently, this last combination corresponds to Veselago's materials. Thus, in this work we focus on the interesting possibility if non-linearities can induce any of the above combinations, excepting the first one. In other words, one of our goals is to understand what might be another observational signature of nonlinearities.

Another of our goals is devoted to study the stability of the above radiation scenario in the presence of other nonlinear electrodynamics.
Of special interest will be to check the effects of the ''medium'' on the production of this radiation. To do this, we will work out the radiated energy for logarithmic electrodynamics following the conventional path of calculating the Poynting vector. As we will see, our analysis renders manifest the vital role played by vacuum electromagnetic nonlinearities in triggering the radiated energy for logarithmic electrodynamics.

Our work is organized as follows. In Sec. 2, we describe the vital role played by nonlinearities in triggering a material with  $(\varepsilon > 0, \mu <0)$, which corresponds to a ferrimagnetic material. Subsequently, the permittivity and susceptibility induced by nonlinearities are considered in order to obtain the refractive index via the dispersion relation for logarithmic electrodynamics.
This would not only provide the theoretical setup for our subsequent work, but also fix the notation. In Sec. 3, we consider the calculation of the electromagnetic radiation. A summary of our work is the subject of Sec. 4.

In our conventions the signature of the metric is ($+1,-1,-1,-1$).

\section{General aspects}
\subsection{On ferrimagnetic materials}

As already mentioned, we now explore the interesting possibility if non-linearities can induce any of the above combinations, excepting the first one. 

For this purpose we begin by considering a generic Lagrangian density:
\begin{equation}
{\cal L} = {\cal L}({\cal F},{\cal G}), \label{Meta05}
\end{equation}
where the arguments of ${\cal L}$ are the usual electromagnetic field invariants, that is, ${\cal F} \equiv  - \frac{1}{4}{F_{\mu \nu }}{F^{\mu \nu }} = \frac{1}{2}\left( {\frac{{{{\bf E}^2}}}{{{c^2}}} - {{\bf B}^2}} \right)$ and 
${\cal G} \equiv  - \frac{1}{4}{F_{\mu \nu }}{\tilde F^{\mu \nu }} = \frac{{\bf E}}{c} \cdot {\bf B}$.\\

Next, after splitting ${F^{\mu \nu }}$ in the sum of a classical background, $F_B^{\mu \nu }$, and a small fluctuation, ${f^{\mu \nu }}$, the corresponding linearized field equations read 
\begin{eqnarray}
{\partial _\mu }\left( {{C_1}{f^{\mu \nu }} + {C_2}{{\tilde f}^{\mu \nu }}} \right) - \frac{1}{2}{\partial _\mu }\left( {k_B^{\mu \nu \kappa \lambda }{f_{\kappa \lambda }} + t_B^{\mu \nu \kappa \lambda }{{\tilde f}_{\kappa \lambda }}} \right) \nonumber\\
- \frac{1}{4}{\partial _\mu }\left( {{\varepsilon ^{\mu \nu \kappa \lambda }}{t_{B\kappa \lambda \rho \sigma }}{f^{\rho \sigma }}} \right)
 =  - {\partial _\mu }\left( {{C_1}F_B^{\mu \nu } + {C_2}{{\tilde F}^{\mu \nu }}} \right) + {j^\nu }, \nonumber\\
 \label{Meta10}
\end{eqnarray}
where $k_B^{\mu \nu \kappa \lambda } = {D_1}F_B^{\mu \nu }F_B^{\kappa \lambda } + {D_2}\tilde F_B^{\mu \nu }\tilde F_B^{\kappa \lambda }$ and $t_B^{\mu \nu \kappa \lambda } = {D_3}F_B^{\mu \nu }F_B^{\kappa \lambda }$. 

Whereas ${C_1} = {\left. {\frac{{\partial {\cal L}}}{{\partial {\cal F}}}} \right|_B}$, ${C_2} = {\left. {\frac{{\partial L}}{{\partial {\cal G}}}} \right|_B}$, ${D_1} = {\left. {\frac{{{\partial ^2}{\cal L}}}{{\partial {{\cal F}^2}}}} \right|_B}$, ${D_2} = {\left. {\frac{{{\partial ^2}{\cal L}}}{{\partial {{\cal G}^2}}}} \right|_B}$ and ${D_3} = {\left. {\frac{{{\partial ^2}{\cal L}}}{{\partial {\cal F}\partial {\cal G}}}} \right|_B}$.

At this stage, we are not bound to consider a constant and uniform electromagnetic background, so that the coefficients $C_1$, $C_2$, $D_1$, $D_2$ and $D_3$ are, in principle, space-time-dependent. This is why the
field equations take the form of eq. (\ref{Meta10}). However, in what follows, we adopt space-time constancy of the background, so that the coefficients above are not acted upon by the space-time derivatives,
giving rise to the constitutive tensors to be present soon below.

However, in what follows we will write the equations of motion in the case $j^{\nu}=0$, in the presence of a constant background with both electric and magnetic fields $({\bf E},{\bf B})$. We thus find
\begin{equation}
\nabla  \cdot {\bf d} = 0, \label{Meta15}
\end{equation}
where ${d_i} = {\varepsilon _{ij}}{e_j} + {\xi _{ij}}{b_j}$. 

Whereas, ${\varepsilon _{ij}}$ and ${\xi _{ij}}$ are given by
\begin{equation}
{\varepsilon _{ij}} = {\delta _{ij}} + {\alpha _i}{E_j} + {\beta _i}{B_j}, \label{Meta20a}
\end{equation}
and
\begin{equation}
{\xi _{ij}} =  - {c^2}{\alpha _i}{B_j} + {\beta _i}{E_j}, \label{Meta20b}
\end{equation}
here we have used the notation $ \boldsymbol{\alpha}  \equiv \frac{1}{{{C_1}}}\left( {\frac{{{D_1}}}{{{c^2}}}\,{\bf E} + \frac{{{D_3}}}{c}\,{\bf B}} \right)$ and $\boldsymbol{\beta}  \equiv \frac{1}{{{C_1}}}\left( {{D_2} \,{\bf B} + \frac{{{D_3}}}{c}\, {\bf E}} \right)$. Note that the tensors $\varepsilon$ and $\xi$ are completely determined by the electromagnetic background. Throughout, ${\bf e}$ and ${\bf b}$ are the electric and magnetic fields arising from the fluctuation ${f^{\mu \nu }}$.

On can now further observe that 
\begin{equation}
\nabla  \times {\bf h} = \frac{1}{{{c^2}}}\frac{\partial }{{\partial t}}{\bf e}, \label{Meta25}
\end{equation}
where ${h_i} \equiv \mu _{ij}^{ - 1}{b_j} + {\eta _{ij}}{e_j}$.

In this case, $\mu _{ij}^{ - 1}$ and ${\eta _{ij}}$ are given by
\begin{equation}
\mu _{ij}^{ - 1} \equiv {\delta _{ij}} - {B_i}{\gamma _j} - {E_i}{\Delta _j}, \label{Meta30a}
\end{equation}
and
\begin{equation}
{\eta _{ij}} \equiv {B_i}{\alpha _j} - \frac{1}{{{c^2}}}{E_i}{\beta _j} =  - \frac{1}{{{c^2}}}{\xi _{ji}}, \label{Meta30b}
\end{equation}
where we have defined $\boldsymbol{\gamma}  \equiv \frac{1}{{{C_1}}}\left( {{D_1}\,{\bf B} - {D_3} \,{\bf E}} \right)$ and 
$\boldsymbol{\Delta}  \equiv \frac{1}{{{C_1}}}\left( { - \frac{{{D_3}}}{c} \,{\bf B} + \frac{{{D_2}}}{{{c^2}}}\, {\bf E}} \right)$.

Incidentally, we would like to point out that, in a recent paper \cite{Casana}, the authors carry out a detailed study of bi-isotropic and bi-anisotropic   material media in terms of a general class of constitutive tensors. In our   case, we stress that it is the vacuum - subject to strong external  electromagnetic fields - that acts as the material medium, with
 permittivity and permeability tensors completely determined by
 the electromagnetic background, according to the equations cast previously. In the particular cases of a purely electric or a purely magnetic external  field, the constitutive tensors naturally arise as symmetric $3\times3$-matrices.

One can easily verify that for an external field ${\bf B}$ (${\bf E}=0$) we have ${\alpha _i} = \frac{{{D_3}}}{{c\,{C_1}}}{B_i}$, ${\beta _i} = \frac{{{D_2}}}{{{C_1}}}{B_i}$, ${\gamma _i} = \frac{{{D_1}}}{{{C_1}}}{B_i}$ and ${\Delta _i} =  - \frac{{{D_3}}}{{c\,{C_1}}}{B_i} =  - {\alpha _i}$. We thus find that, 
\begin{equation}
{\varepsilon _{ij}} = {\delta _{ij}} + \frac{{{D_2}}}{{{C_1}}}{B_i}{B_j}, \label{Meta35a}\end{equation}
\begin{equation}
{\xi _{ij}} =  - c\,\frac{{{D_3}}}{{{C_1}}}{B_i}{B_j}, \label{Meta35b}
\end{equation}
\begin{equation}
\mu _{ij}^{ - 1} = {\delta _{ij}} - \frac{{{D_1}}}{{{C_1}}}{B_i}{B_j}, \label{Meta35c}\end{equation}
and
\begin{equation}
{\eta _{ij}} = \frac{{{D_3}}}{{c\,{C_1}}}{B_i}{B_j} =  - \frac{1}{{{c^2}}}{\xi _{ij}}.\label{Meta35d}
\end{equation}

By defining the matrix ${{\cal B}_{ij}} \equiv {B_i}{B_j}$, which is a symmetric one with eigenvalues $0,0$ and ${{\bf B}^2}$, the expressions for $d$ and $h$, in matrix notation, become
\begin{equation}
d = e + \frac{{{D_2}}}{{{C_1}}}{\cal B}\,e - c\,\frac{{{D_3}}}{{{C_1}}}{\cal B}\,b, \label{Meta40a}
\end{equation}
\begin{equation}
h = b - \frac{{{D_1}}}{{{C_1}}}{\cal B}\,b + \frac{{{D_3}}}{{c\,{C_1}}}{\cal B}\,e. \label{Meta40b}
\end{equation}
Whereas
\begin{equation}
\nabla  \cdot {\bf d} = 0, \label{Meta40c}
\end{equation}
\begin{equation}
\nabla  \times {\bf h} = \frac{1}{{{c^2}}}\frac{\partial }{{\partial t}}{\bf d}. \label{Meta40d}
\end{equation}

After the matrix ${\cal B}$ is diagonalized, we then have
\begin{equation}
{\partial _i}{d_i} = 0, \label{Meta45}
\end{equation}
\begin{equation}
{\varepsilon _{ijk}}{\partial _j}{h_k} = \frac{1}{{{c^2}}}\frac{\partial }{{\partial t}}{d_i}, \label{Meta50}
\end{equation}
where the matrix ${\cal B} = \left( {\begin{array}{*{20}{c}}
{{{\bf B}^2}}&0&0\\
0&0&0\\
0&0&0
\end{array}} \right)$. \\

An immediate consequence of this result is that 
${d_x} = \left( {1 + \frac{{{D_2}}}{{{C_1}}}{{\bf B}^2}} \right){e_x} - c\,\frac{{{D_3}}}{{{C_1}}}{{\bf B}^2}\,{b_x}$, ${d_y} = {e_y}$ and ${d_z} = {e_z}$. We also find that  ${h_x} = \left( {1 - \frac{{{D_1}}}{{{C_1}}}{{\bf B}^2}} \right){b_x} + \frac{{{D_3}}}{{c\,{C_1}}}{{\bf B}^2}\, {e_x}$, ${h_y} = {b_y}$ and ${h_z} = {b_z}$.\\

This leads to the following expressions for ${\varepsilon _{xx}}$ and ${\mu _{xx}}$, that is,
\begin{equation}
{\varepsilon _{xx}} = 1 + \frac{{{D_2}}}{{{C_1}}}{{\bf B}^2}, \label{Meta55}
\end{equation}
and
\begin{equation}
\mu _{xx}^{ - 1} = 1 - \frac{{{D_1}}}{{{C_1}}}{{\bf B}^2}. \label{Meta60}
\end{equation}
It should be further noted that, in principle, the positivity of the above expressions is not assured.
However, from equation (\ref{Meta60}) it is evident that
\begin{equation}
{\mu _{xx}} = \frac{1}{{1 - \frac{{{D_1}}}{{{C_1}}}{{\bf B}^2}}},
\end{equation}
this then implies that ${\mu _{xx}}<0$ if $\frac{{{D_1}}}{{{C_1}}}{{\bf B}^2} > 1$. We see,  therefore, a remarkable feature of nonlinearities of the electromagnetic field.

The preceding considerations clearly show that the nonlinearities induce the second combination mentioned above $(\varepsilon > 0, \mu <0)$, which corresponds to a ferrimagnetic material in the classification given in  \cite{Caloz}.

So far our treatment is completely general. However, for the specific case of logarithmic electrodynamics, the permittivity and susceptibility induced by nonlinearities will be  considered in the next Subsection.

\subsection{Some features of logarithmic electrodynamics} 

We now proceed to explore other relevant aspects on nonlinearities. Let us commence our undertaking by considering logarithmic electrodynamics \cite{Logarithmic}: 
\begin{equation}
{\cal L} =  - \beta ^2 \ln \left[ {1 - \frac{{{\cal F}}}{{\beta ^2 }} - \frac{{{\cal G}^2 }}{{2\beta ^4 }}} \right],
\label{log05}
\end{equation}
recalling again that ${\cal F} = \frac{1}{2}\left( {\frac{{{{\bf E}^2}}}{{{c^2}}} - {{\bf B}^2}} \right)$ and ${\cal G} = \frac{{\bf E}}{c} \cdot {\bf B}$. We further note that the parameter $\beta$ measures the nonlinearity of the theory and in the limit $\beta \to \infty$ the Lagrangian (\ref{log05}) reduces to the Maxwell theory.

As already stated, in this Subsection we will be mainly interested in the  dispersion relations for the electrodynamics under consideration. The first step in this direction is to consider a generic Lagrangian density ${\cal L}={\cal L}({\cal F},{\cal G})$, in the presence of a constant background with both $({\bf E},{\bf B})$. As before, after splitting $F^{\mu\nu}$ in the sum of a classical background $F^{\mu\nu}_{B}$, and a small fluctuation, $f^{\mu\nu}$, the corresponding linearized equations of motion read
\begin{equation}
\nabla  \cdot {\bf d} = {\partial _i}{d_i} = {\varepsilon _{ij}}{\partial _i}{e_j} + {\xi _{ij}}{\partial _i}{b_j} = 0, \nonumber\\
\end{equation}
\begin{equation}
\nabla  \times {\bf e} =  - \frac{\partial }{{\partial t}}{\bf b}\,\,\, \Rightarrow {\varepsilon _{ijk}}{\partial _j}{e_k} =  - \frac{\partial }{{\partial t}}{b_i}, \nonumber\\
\end{equation}
\begin{equation}
\nabla  \cdot {\bf b} = 0,\nonumber\\
\end{equation}
\begin{eqnarray}
\nabla  \times {\bf h} = \frac{1}{{{c^2}}}\frac{\partial }{{\partial t}}{\bf e} \nonumber\\
\Rightarrow{\varepsilon _{ijk}}\mu _{kl}^{ - 1}{\partial _j}{b_l} + {\varepsilon _{ijk}}{\eta _{kl}}{\partial _j}{e_l} &=& \frac{1}{{{c^2}}}{\varepsilon _{ij}}{\partial _t}{e_j} + \frac{1}{{{c^2}}}{\xi _{ij}}{\partial _t}{b_j}. \nonumber\\
\label{log10}
\end{eqnarray}
Throughout, ${\bf e}$ and ${\bf b}$ are the electric and magnetic fields arising from the fluctuation $f^{\mu\nu}$.

By considering the plane waves
\begin{equation}
e_{i} = e_{0i}\,{e^{i\left( {{\bf k} \cdot {\bf x} - wt} \right)}}, \,\,\,\,\, b_{i} = b_{0i}\,{e^{i\left( {{\bf k} \cdot {\bf x} - wt} \right)}}, \label{log15}
\end{equation}
from the Eqs. (\ref{log10}) it follows that
\begin{equation}
 {M_{in}}\left( {w,{\bf k};{{\bf E}_B},{{\bf B}_B}} \right){e_{0n}} = 0, \label{log20}
\end{equation}
where 
\begin{eqnarray}
 {M_{in}} &=& \frac{{{w^2}}}{{{c^2}}}{\varepsilon _{in}} + \frac{w}{{{c^2}}}{\xi _{ij}}{\varepsilon _{jmn}}{k_m} + {\varepsilon _{ijk}}{\varepsilon _{lmn}}\mu _{kl}^{ - 1}{k_j}{k_m}  \nonumber\\
 &+& w{\varepsilon _{ijk}}{\eta _{kn}}{k_j},\label{log25}
\end{eqnarray}
in which we have used
\begin{equation}
{\varepsilon _{ij}} = {\delta _{ij}} + {\alpha _i}{E_j} + {\beta _i}{B_j}, \label{log30a}
\end{equation}
\begin{equation}
{\xi _{ij}} =  - {c^2}{\alpha _i}{B_j} + {\beta _i}{E_j}, \label{log30b}
\end{equation}
\begin{equation}
\mu _{ij}^{ - 1} = {\delta _{ij}} - {B_i}{\gamma _j} - {E_i}{\Delta _j}, \label{log30c}
\end{equation}
\begin{equation}
{\eta _{ij}} =  - \frac{1}{{{c^2}}}{\xi _{ji}} = {\alpha _j}{B_i} - \frac{1}{{{c^2}}}{\beta _j}{E_i}. \label{log30d}
\end{equation}
In passing we recall that ${\alpha _i}$, ${\beta _i}$, ${\gamma _i}$ and ${\Delta _i}$ are given in terms of $C_{1}$, $D_{1}$, $D_{2}$ and $D_{3}$.

It is of interest to note that in the particular case of an external electric field ${\bf E}=0$, Eq. (\ref{log20}) takes the particularly simple form
\begin{equation}
\left( {\frac{{{w^2}}}{{{c^2}}}{\varepsilon _{in}} + {\varepsilon _{ijk}}{\varepsilon _{lmn}}\mu _{kl}^{ - 1}{k_j}{k_m}} \right){e_{0n}} = 0.  \label{log35}
\end{equation}
Here we have used that $D_{3}=0$ (${{\cal G}_{Background}} = 0$), whereas ${\xi _{ij}} = 0$ and ${\eta _{ij}} = 0$. We accordingly express Eq. (\ref{log35}) in the form
\begin{eqnarray}
&\biggl[&{{\!\!\!\! \frac{{{w^2}}}{{{c^2}}}{\varepsilon _{in}} + \mu _{in}^{ - 1}{{\bf k}^2} + \left( {tr{\mu ^{ - 1}}} \right){k_i}{k_n} + {\delta _{in}}\left( {\mu _{jk}^{ - 1}{k_j}{k_k}} \right)}}  \nonumber\\
 &-&\!\! {{k_i}\mu _{nj}^{ - 1}{k_j} - \mu _{ij}^{ - 1}{k_j}{k_n} - {\delta _{in}}\left( {tr{\mu ^{ - 1}}} \right){{\bf k}^2}} \biggr]{e_{0n}} = 0,  \label{log40}
\end{eqnarray}
where
\begin{equation}
{\varepsilon _{in}} = {\delta _{in}} + \frac{{{D_2}}}{{{C_1}}}{B_i}{B_n} \equiv {\varepsilon _{ni}}, \label{log45a}
\end{equation}
\begin{equation}
\mu _{jk}^{ - 1} = {\delta _{jk}} - \frac{{{D_1}}}{{{C_1}}}{B_j}{B_k} \equiv \mu _{kj}^{ - 1}, \label{log45b}
\end{equation}
\begin{equation}
tr{\mu ^{ - 1}} = 3 - \frac{{{D_1}}}{{{C_1}}}{{\bf B}^2}. \label{log45c}
\end{equation}

Making use of these relations, we find that Eq. (\ref{log40}) reduces to
\begin{eqnarray}
\left[ {\frac{{{w^2}}}{{{c^2}}}{\mu _{mi}}{\varepsilon _{in}} + {{\bf k}^2}{\delta _{mn}} - {k_m}{k_n} + \left( {tr{\mu ^{ - 1}}} \right){k_i}{\mu _{im}}k_{n}} \right. \nonumber\\
\left. { - \left( {tr{\mu ^{ - 1}}} \right){{\bf k}^2}{\mu _{mn}} + \left( {\mu _{ij}^{ - 1}{k_i}{k_j}} \right){\mu _{mn}} - {k_i}{k_j}\mu _{jn}^{ - 1}} \right]{e_{0n}} = 0, \nonumber\\
\label{log50}
\end{eqnarray}
where  ${\mu _{ij}} = {{\delta _{ij}} + \frac{{{{{D_1}} \mathord{\left/
 {\vphantom {{{D_1}} {{C_1}}}} \right.
 \kern-\nulldelimiterspace} {{C_1}}}}}{{1 - {{{D_1}} \mathord{\left/
 {\vphantom {{{D_1}} {{C_1}{B^2}}}} \right.
 \kern-\nulldelimiterspace} {{C_1}{{\bf B}^2}}}}}{B_i}{B_j}}$. \\

 It is also important to observe that in the configuration space we have $w = i{\partial _t}$, ${k_i} =  - i{\partial _i}$ and ${{\bf k}^2} =  - {\nabla ^2}$. Hence, we readily verify that Eq. (\ref{log50}) can be brought to the form
\begin{eqnarray}
\left[ {-\frac{1}{{{c^2}}}\partial _t^2{\mu _{mi}}{\varepsilon _{in}} - {\delta _{mn}}{\nabla ^2} + {\partial _m}{\partial _n} - \left( {tr{\mu ^{ - 1}}} \right){\mu _{mi}}{\partial _i}{\partial _n}} \right. + \nonumber\\
\left. {  \left( {tr{\mu ^{ - 1}}} \right){\mu _{mn}}{\nabla ^2} - {\mu _{mn}}\mu _{ij}^{ - 1}{\partial _i}{\partial _j} + {\mu _{mi}}\mu _{nj}^{ - 1}{\partial _i}{\partial _j}} \right]{e_{0n}} = 0. \nonumber\\
\label{log55}
\end{eqnarray}

Thus, finally we end up with
\begin{eqnarray}
\left[ {{\mu _{mi}}{\varepsilon _{in}}\frac{1}{{{c^2}}}\partial _t^2 - \left( {{\mu _{mn}}tr{\mu ^{ - 1}} - {\delta _{mn}}} \right){\nabla ^2} - {\partial _m}{\partial _n}} \right. + \nonumber\\
\left. { \left( {tr{\mu ^{ - 1}}} \right){\mu _{mi}}{\partial _i}{\partial _n} + \left( {{\mu _{mn}}\mu _{ij}^{ - 1} - {\mu _{mi}}\mu _{nj}^{ - 1}} \right){\partial _i}{\partial _j}} \right]{e_{0n}} = 0. \nonumber\\
\label{log60}
\end{eqnarray}

Before proceeding our analysis of the dispersion relation, we call attention to the fact that
\begin{eqnarray}
{C_1} = \frac{{\partial {\cal L}}}{{\partial {\cal F}}} = \frac{1}{{1 - \frac{{\cal F}}{{{\beta ^2}}} - \frac{{{{\cal G}^2}}}{{2{\beta ^4}}}}}, \nonumber\\
{C_2} = \frac{{\partial {\cal L}}}{{\partial {\cal G}}} = \frac{1}{{{\beta ^2}}}{\cal G}{C_1}, \nonumber\\
{D_1} = \frac{{{\partial ^2}{\cal L}}}{{\partial {{\cal F}^2}}} = \frac{1}{{{\beta ^2}}}C_1^2, \nonumber\\
{D_2} = \frac{{{\partial ^2}{\cal L}}}{{\partial {{\cal G}^2}}} = \frac{1}{{{\beta ^2}}}{C_1} + \frac{1}{{{\beta ^6}}}{{\cal G}^2}C_1^2, \nonumber\\
{D_3} = \frac{{{\partial ^2}{\cal L}}}{{\partial {\cal F}\partial {\cal G}}} = \frac{1}{{{\beta ^4}}}{\cal G}C_1^2. \label{log65}
\end{eqnarray}

Restricting our considerations to the ${\bf E}=0$ case, we have ${\cal F} =  - \frac{1}{2}{{\bf B}^2}$ and ${\cal G}=0$. We thus find ${C_1} = \frac{1}{{1 + \frac{{{{\bf B}^2}}}{{2{\beta ^2}}}}}$, $C_{2}=0$, $\frac{{{D_1}}}{{{C_1}}} = \frac{1}{{{\beta ^2} + \frac{{{{\bf B}^2}}}{2}}}$, $\frac{{{D_2}}}{{{C_1}}} = \frac{1}{{{\beta ^2}}}$ and $D_{3}=0$. Making use of the foregoing results one encounters that, ${\varepsilon _{ij}} = {\delta _{ij}} + \frac{{{D_2}}}{{{C_1}}}{B_i}{B_j}$, have two eigenvalues 1 and $1 + \frac{{{D_2}}}{{{C_1}}}{{\bf B}^2}$. In fact, for logarithmic electrodynamics this eigenvalue reduces to $1 + \frac{{{{\bf B}^2}}}{{{\beta ^2}}}$. Similarly, from ${\mu _{ij}} = {\delta _{ij}} + \frac{{\frac{{{D_1}}}{{{C_1}}}}}{{1 - \frac{{{D_1}}}{{{C_1}}}{{\bf B}^2}}}{B_i}{B_j}$, we again have two eigenvalues $1$ and $\frac{1}{{1 - \frac{{{D_1}}}{{{C_1}}}{{\bf B}^2}}}$. For logarithmic electrodynamics the previous eigenvalue becomes $\frac{{{\beta ^2} + \frac{{{{\bf B}^2}}}{2}}}{{{\beta ^2} - \frac{{{{\bf B}^2}}}{2}}}$.

Next, by making use of ${\varepsilon _{in}}$, $\mu _{in}^{ - 1}$ and $tr{\mu ^{ - 1}}$ in the dispersion matrix ($M_{in}$), we can write the corresponding matrix as:
\begin{eqnarray}
{M_{in}} &=& \left( {\frac{{{w^2}}}{{{c^2}}} - {{\bf k}^2}} \right){\delta _{in}} + \frac{{{D_1}}}{{{C_1}}}\left( {{{\bf B}^2}{{\bf k}^2} - {{\left( {{\bf B} \cdot {\bf k}} \right)}^2}} \right){\delta _{in}} \nonumber\\
&+& \left( {\frac{{{D_2}}}{{{C_1}}}\frac{{{w^2}}}{{{c^2}}} - \frac{{{D_1}}}{{{C_1}}}{{\bf k}^2}} \right){B_i}{B_n} + \left( {1 - \frac{{{D_1}}}{{{C_1}}}{{\bf B}^2}} \right){k_i}{k_n} \nonumber\\
&+& \frac{{{D_1}}}{{{C_1}}}\left( {{\bf B} \cdot {\bf k}} \right)\left( {{k_i}{B_n} + {k_n}{B_i}} \right).  \label{log70}
\end{eqnarray}
We are now in position to examine the condition, $\det M = 0$, in order to obtain the dispersion relations. We also recall that the index of refraction is given by $n \equiv \frac{{|{\bf k}|c}}{w}$, and after some manipulations, it follows that
\begin{eqnarray}
M_{in}&=&\left[ {\frac{1}{{{n^2}}} - 1 + \frac{{{D_1}}}{{{C_1}}}\left( {{{\bf B}^2} - {{\left( {{\bf B} \cdot \hat {\bf k}} \right)}^2}} \right)} \right]{\delta _{in}} \nonumber\\
&+& \left( {\frac{{{D_2}}}{{{C_1}}}\frac{1}{{{n^2}}} - \frac{{{D_1}}}{{{C_1}}}} \right){B_i}{B_n} \nonumber\\
&+&\left( {1 - \frac{{{D_1}}}{{{C_1}}}{{\bf B}^2}} \right){{\hat {\bf k}}_i}{{\hat {\bf k}}_n} + \frac{{{D_1}}}{{{C_1}}}\left( {{\bf B} \cdot \hat {\bf k}} \right)\left( {{{\hat k}_i}{B_n} + {{\hat k}_n}{B_i}} \right), \nonumber\\
\label{log75}
\end{eqnarray}
where $\hat {\bf k} \equiv \frac{{\bf k}}{{|{\bf k}|}}$.

This last expression clearly shows that the $M$-matrix does not depend on $|{\bf k}| = {{2\pi } \mathord{\left/
 {\vphantom {{2\pi } \lambda }} \right.\kern-\nulldelimiterspace} \lambda }$. Then the refractive index, $n$, arising from the condition, $det M=0$, does not depend on $\lambda$ but of the relative direction between the propagation vector $\hat {\bf k}$ and the external field. In this manner, we obtain an effective refractive index $n = n\left( {{\bf B},\hat {\bf k}} \right)$. 

To further elaborate on the comparative features of the index of refraction, we shall examine two different situations. First, we consider ${n_ \bot }$ if $\hat {\bf k} \,\bot \,{\bf B}$. In this case, the condition, $det M=0$, reads
\begin{eqnarray}
\det \left[ {\left( {\frac{1}{{{n^2}}} - 1 + \frac{{{D_1}}}{{{C_1}}}{{\bf B}^2}} \right)} \right.{\delta _{in}} + \left( {1 - \frac{{{D_1}}}{{{C_1}}}{{\bf B}^2}} \right){{\hat k}_i}{{\hat k}_n} 
\nonumber\\
\left. {\left( {\frac{{{D_2}}}{{{C_1}}}\frac{1}{{{n^2}}} - \frac{{{D_1}}}{{{C_1}}}} \right){B_i}{B_n}} \right] = 0. \label{log80}
\end{eqnarray}
From this equation it is clear that the determinant has the form $\det \, \left( {a{\delta _{ij}} + b{u_i}{u_j} + c{v_i}{v_j}} \right)$, whose solution is given by $a\left[ {\left( {a + b{u^2}} \right)\left( {a + c{v^2}} \right) - bc{{\left( {u \cdot v} \right)}^2}} \right]$. It is a simple matter to verify that the condition, $det M=0$, becomes
\begin{equation}
\frac{1}{{{n^2}}}\left( {\frac{1}{{{n^2}}} - 1 + \frac{{{D_1}}}{{{C_1}}}{{\bf B}^2}} \right)\left[ {\frac{1}{{{n^2}}}\left( {1 + \frac{{{D_2}}}{{{C_1}}}{{\bf B}^2}} \right) - 1} \right] = 0. \label{log85}
\end{equation}
Thus, we finally obtain two modes associated to the direction of propagation $\hat {\bf k}$, that is,
\begin{equation}
n_ \bot ^2 = \frac{1}{{1 - \frac{{{D_1}}}{{{C_1}}}{{\bf B}^2}}}, \label{log90}
\end{equation}
and
\begin{equation}
n_ \bot ^2 = 1 + \frac{{{D_2}}}{{{C_1}}}{{\bf B}^2}. \label{log90}
\end{equation}
For logarithmic electrodynamics we have $\frac{{{D_1}}}{{{C_1}}} = \frac{{{C_1}}}{{{\beta ^2}}} = \frac{1}{{{\beta ^2} + \frac{{{{\bf B}^2}}}{2}}}$ and $\frac{{{D_2}}}{{{C_1}}} =  - \frac{1}{{{\beta ^2}}}$. Hence we see that the two modes take the form 
\begin{equation}
n_ \bot ^2 = \frac{{{\beta ^2} + \frac{{{{\bf B}^2}}}{2}}}{{{\beta ^2} - \frac{{{{\bf B}^2}}}{2}}},\label{log95}
\end{equation}
and
\begin{equation}
n_ \bot ^2 = 1 + \frac{{{{\bf B}^2}}}{{{\beta ^2}}}. \label{log100}
\end{equation}

Second, we consider ${n_\parallel }$ if ${\hat {\bf k}\,\parallel \, {\bf B}}$. By using ${\bf B} = \xi |{\bf B}|\hat {\bf k}$ and ${\bf B} \cdot \hat {\bf k} = \xi |{\bf B}|$, where $\xi  =  \pm 1$ stands parallel or anti-parallel to the propagation direction. After some manipulations the condition $det M=0$ becomes
\begin{equation}
\frac{1}{{{n^2}}}{\left( {\frac{1}{{{n^2}}} - 1} \right)^2}\left( {1 + \frac{{{D_2}}}{{{C_1}}}{{\bf B}^2}} \right) = 0.\label{log105}
\end{equation}
In this case, the corresponding mode associated to the direction of propagation $\hat {\bf k}$ becomes ${n_\parallel } = 1$.

In summary then, we easily verify that the previous electromagnetic vacuum acts like a birefringent medium with two indices of refraction determined by the relative direction between the propagation vector $\hat {\bf k}$ and the external field. More recently, this has also helped us to gain insights into the peculiarities about vacuum nonlinearities such as calculating the bending of light \cite{inverse}.

\section{Electromagnetic radiation}

As already mentioned, our immediate objective is to compute the electromagnetic radiation produced by a moving charged particle interacting with  a medium characterized by nonlinearities of the electromagnetic field.
With this in view, the starting point are the Maxwell equations for a moving charged particle in a medium characterized by logarithmic electrodynamics:
\begin{eqnarray}
\nabla  \cdot {\bf e} = \frac{{4\pi }}{\varepsilon }{\rho _{ext}}, \nonumber\\
\nabla  \cdot {\bf b} = 0, \nonumber\\
\nabla  \times {\bf e} =  - \frac{1}{c} \frac{{\partial {\bf b}}}{{\partial t}}, \nonumber\\
\nabla  \times {\bf b} = \frac{{\varepsilon \mu }}{c}\frac{{\partial {\bf e}}}{{\partial t}} + \frac{{4\pi \mu }}{c}{{\bf j}_{ext}}. \label{Red65}
\end{eqnarray}
where ${\rho _{ext}}$ and ${{\bf j}_{ext}}$ denote the external charge and current densities. Whereas ${\bf d} = \varepsilon {\bf e}$ and ${\bf b} = \mu {\bf h}$. Here we have simplified our notation by setting ${\bf E}_{p}={\bf e}$ and ${\bf B}_{p}={\bf b}$.

It is straightforward to see that the foregoing equations can be written alternatively in the form
\begin{equation}
{\nabla ^2}{\bf b} - \frac{{\varepsilon \mu }}{{{c^2}}}\frac{{{\partial ^2}{\bf b}}}{{\partial {t^2}}} =  - \frac{{4\pi \mu }}{c}\nabla  \times {{\bf j}_{ext}}, \label{Red70}
\end{equation}
and
\begin{equation}
{\nabla ^2}{\bf e} - \frac{{\varepsilon \mu }}{{{c^2}}}\frac{{{\partial ^2}{\bf e}}}{{\partial {t^2}}} = \frac{{4\pi \mu }}{{{c^2}}}\frac{{\partial {{\bf j}_{ext}}}}{{\partial t}} + \frac{{4\pi }}{\varepsilon }\nabla {\rho _{ext}}, \label{Red75}
\end{equation}
where the external charge and current densities are given by: ${\rho _{ext}}\left( {t,{\bf x}} \right) = Q\delta \left( x \right)\delta \left( y \right)\delta \left( {z - vt} \right)$ and ${\bf j}\left( {t,{\bf x}} \right) = Qv\delta \left( x \right)\delta \left( y \right)\delta \left( {z - vt} \right){\hat {\bf e}_z}$.
In passing we note that, for simplicity, we are considering the $z$ axis as the direction of the moving charged particle. 

Next, in order to solve equations (\ref{Red70}) and (\ref{Red75}), we shall begin by performing a Fourier transform to momentum space via
\begin{equation}
f(t,{\bf x}) = \int {\frac{{dw{d^3}{\bf k}}}{{{{\left( {2\pi } \right)}^4}}}} {e^{ - iwt + {\bf k} \cdot {\bf x}}}f\left( {w,{\bf k}} \right), \label{Red80}
\end{equation}
where $f$ stands for the electric and magnetic fields. Then, the corresponding electric and magnetic fields read:
\begin{equation}
{\bf b}\left( {w,{\bf k}} \right) = - \,i\,\frac{{4\pi \mu }}{c}\frac{{{\bf k} \times {{\bf j}_{ext}}\left( {w,{\bf k}} \right)}}{\cal O}, \label{Red85}
\end{equation}
and
\begin{equation}
{\bf e}\left( {w,{\bf k}} \right) = i\,\frac{{4\pi }}{\varepsilon }\frac{{{\bf k}\,{\rho _{ext}}\left( {w,{\bf k}} \right)}}{\cal O} - i\frac{{4\pi \mu w}}{{{c^2}}}\frac{{{{\bf j}_{ext}}\left( {w,{\bf k}} \right)}}{\cal O},\label{Red90}
\end{equation}
where 
\begin{equation}
{\cal O} = \frac{{{w^2}}}{{{c^{\prime 2}}}} - {{\bf k}^2},  \  \  \
\frac{1}{{{c^{\prime 2}}}} \equiv \frac{{\varepsilon \mu }}{{{c^2}}}. \label{Red91}
\end{equation}
In the same way, the external charge and current densities, in the Fourier space, take the form: $\rho_{ext} \left( {w,{\bf k}} \right) = 2\pi Q\delta \left( {w - {k_z}v} \right)$ and ${{\bf j}_{ext}}\left( {w,{\bf k}} \right) = 2\pi Qv\delta \left( {w - {k_z}v} \right){\hat {\bf e}_z}$.

From the above we can proceed to obtain ${\bf b}\left( {w,{\bf x}} \right)$ and ${\bf e}\left( {w,{\bf x}} \right)$. It is clear now that ${\bf b}\left( {w,{\bf x}} \right)$ is given by
\begin{equation} 
{\bf b}\left( {w,{\bf x}} \right) = \int {\frac{{{d^3}{\bf k}}}{{{{\left( {2\pi } \right)}^3}}}} \; {e^{i{\bf k} \cdot {\bf x}}}\;{\bf b}\left( {w,{\bf k}} \right). \label{Red95}
\end{equation}

We may now take advantage of the axial symmetry of the problem under consideration. If we take cylindrical coordinates, equation (\ref{Red95}) becomes
\begin{eqnarray}
{\bf b}\left( {w,{\bf x}} \right) &=&  - \frac{{i\mu Qv}}{{\pi c}}{e^{i{{wz} \mathord{\left/
 {\vphantom {{wz} v}} \right.
 \kern-\nulldelimiterspace} v}}}\int_0^\infty  {d{k_T}{k_T}} \nonumber\\
 &\times&\int_0^{2\pi } {d\alpha } \frac{{{e^{i{k_T}{x_T}\cos \alpha }}}}{\left. {\cal O} \right|_{{k_z} = {w \mathord{\left/
 {\vphantom {w v}} \right.
 \kern-\nulldelimiterspace} v}}} \left( {{k_T}\sin \alpha\, \pmb{\hat \rho}  - {k_T}\cos \alpha\, \pmb{\hat \phi} } \right). \nonumber\\
\label{Red100}
\end{eqnarray}
In passing we recall that $\int_0^{2\pi } {d\theta } {e^{ix\cos \theta }}\sin \theta  = 0$ and $\int_0^{2\pi } {d\theta } {e^{ix\cos \theta }}\cos \theta  = 2\pi i{J_1}\left( x \right)$, which implies
\begin{equation}
{\bf b}\left( {w,{\bf x}} \right) =  - \frac{{2\mu Qv}}{c}{e^{i{{wz} \mathord{\left/
 {\vphantom {{wz} v}} \right.
 \kern-\nulldelimiterspace} v}}} \int_0^\infty  {d{k_T}\;k_T^2} \;\frac{{{J_1}\left( {{k_T}{x_T}} \right)}}{{{{\left. {\cal O} \right|}_{{k_z} = {w \mathord{\left/
 {\vphantom {w v}} \right.
 \kern-\nulldelimiterspace} v}}}}}\;\pmb{\hat \phi}, \label{Red105}
\end{equation}
where ${{J_1}\left( {{k_T}{x_T}} \right)}$ is a Bessel function of the first kind.

In this case, ${\left. {\cal O} \right|_{{k_z} = {w \mathord{\left/
 {\vphantom {w v}} \right.
 \kern-\nulldelimiterspace} v}}} = {w^2}\left( {\frac{1}{{{c^{\prime 2}}}} - \frac{1}{{{v^2}}}} \right) - {\bf k}_T^2$. From this last expression it follows that
\begin{equation}
{\bf b}\left( {w,{\bf x}} \right) = \frac{{2\mu Qv}}{c}{e^{i{{wz} \mathord{\left/
 {\vphantom {{wz} v}} \right.
 \kern-\nulldelimiterspace} v}}} \int_0^\infty  {d{k_T}}\; k_T^2\;\frac{{{J_1}\left( {{k_T}{x_T}} \right)}}{{\left( {k_T^2 + {\sigma}^{2}} \right)}}\; \pmb {\hat \phi}, \label{Red110}
\end{equation}
where ${\sigma ^2} = {w^2}\left( {\frac{1}{{{v^2}}} - \frac{1}{{{c^{\prime 2}}}}} \right)$.
We next observe that the previous expression can be brought to the form 
\begin{eqnarray}
{\bf b}\left( {w,{\bf x}} \right) &=& \frac{{2\mu Qv}}{c}{e^{i{{wz} \mathord{\left/
 {\vphantom {{wz} v}} \right.
 \kern-\nulldelimiterspace} v}}}\int_0^\infty  {dy\,{e^{-y{\sigma}^{2}}}} \nonumber\\
&\times&\int_0^\infty  {d{k_T}}\; k_T^2\;{e^{ - yk_T^2}}{J_1}\left( {{k_T}{x_T}} \right)\pmb {\hat \phi}.\label{Red115}
\end{eqnarray}
From this last expression it follows that
\begin{equation}
{\bf b}\left( {w,{\bf x}} \right) = \! \frac{{2\mu Qv}}{c}{e^{i{{wz} \mathord{\left/
 {\vphantom {{wz} v}} \right.
 \kern-\nulldelimiterspace} v}}}\, \frac{{{x_T}}}{4}\int_0^\infty  {dy\frac{1}{{{y^2}}}{e^{-y {\sigma}^{2} - {{x_T^2} \mathord{\left/
 {\vphantom {{x_T^2} {4y}}} \right.
 \kern-\nulldelimiterspace} {4y}}}}} \,\pmb {\hat \phi},  \label{Red120}
\end{equation}
or, in terms of the modified Bessel function,
equation (\ref{Red120}), becomes
\begin{equation}
{\bf b}\left( {w,{\bf x}} \right) = \frac{{2\mu Qv}}{c}\,{e^{i{{wz} \mathord{\left/
{\vphantom {{wz} v}} \right.
\kern-\nulldelimiterspace} v}}}\,\sigma {K_1}\left( {\sigma {x_T}} \right)\,\pmb {\hat \phi}, \label{Red125}
\end{equation}
where, in cylindrical coordinates, ${x_T} = \rho$.

Now we come to the calculation of the electric field. From the expression (\ref{Red90}), we find that the electric field may be written in the form 
\begin{eqnarray}
{\bf e}\left( {w,{\bf x}} \right) &=& i\,8 {\pi}^{2} Q\int {\frac{{{d^3}k}}{{{{\left( {2\pi } \right)}^3}}}} \frac{w}{{v\;{\cal O}}}\;\delta \left( {{k_z} - {\raise0.7ex\hbox{$w$} \!\mathord{\left/
 {\vphantom {w v}}\right.\kern-\nulldelimiterspace}
\!\lower0.7ex\hbox{$v$}}} \right){e^{i{\bf k} \cdot {\bf x}}} \nonumber\\
&\times&\left\{ {\frac{{{k_x}}}{{\epsilon }}\,\pmb{{\hat e}_x} + \frac{{{k_y}}}{{\epsilon }}\,\pmb{{\hat e}_y} + \left( {\frac{{{k_z}}}{{\epsilon }} - \frac{\mu\,w\,v}{c^{2}}} \right)\pmb{{\hat e}_z}} \right\}. \nonumber\\
\label{Red130}
\end{eqnarray}

In the same way as was done for the magnetic field, we then get
\begin{eqnarray}
{e_\rho }\left( {w,{\bf x}} \right) &=& \frac{{i\,Q}}{{v{{{\pi }}}}}  \int_0^\infty {d{k_T}\,{k_T}}\,{e^{iw{z \mathord{\left/
 {\vphantom {z v}} \right.
 \kern-\nulldelimiterspace} v}}}\frac{1}{{{{\left. {\cal O} \right|}_{{k_z} = {w \mathord{\left/
 {\vphantom {w v}} \right.
 \kern-\nulldelimiterspace} v}}}}} \nonumber\\
&\times&\left\{ {\frac{{2\pi i{k_T}}}{{\epsilon }}{J_1}\left( {{k_T}{x_T}} \right)} \right\}, \label{Red135}
\end{eqnarray}
and
\begin{eqnarray}
{e_z}\left( {w,{\bf x}} \right) &=& \frac{{i\,Q}}{{v{{{\pi }}}}} \int_0^\infty {d{k_T}\,{k_T}}\,{e^{iw{z \mathord{\left/
 {\vphantom {z v}} \right.
 \kern-\nulldelimiterspace} v}}}\frac{1}{{{{\left. {\cal O} \right|}_{{k_z} = {w \mathord{\left/
 {\vphantom {w v}} \right.
 \kern-\nulldelimiterspace} v}}}}} \nonumber\\
&\times&\left\{ {2\pi\frac{w}{\epsilon}\left( {\frac{1}{{v}} - \frac{v}{c^{\prime2}}} \right){J_0}\left( {{k_T}{x_T}} \right)} \right\}.
\label{Red140}
\end{eqnarray}

The integral occurring on the right-hand side of the previous expressions can be as before. We get accordingly 
\begin{equation}
{e_\rho }\left( {w,{\bf x}} \right) = \frac{{2Q}}{{v\varepsilon }}{e^{i{{wz} \mathord{\left/
 {\vphantom {{wz} v}} \right.
 \kern-\nulldelimiterspace} v}}}\sigma {K_1}\left( {\sigma {x_T}} \right), \label{Red145}
\end{equation}
and
\begin{equation}
 {e_z}\left( {w,{\bf x}} \right) =  -\,i\,\frac{{2Q}}{{v\varepsilon }}{e^{i{{wz} \mathord{\left/
 {\vphantom {{wz} v}} \right.
 \kern-\nulldelimiterspace} v}}}w\left( {\frac{1}{v} - \frac{v}{{{c^{\prime 2}}}}} \right){K_0}\left( {\sigma {x_T}} \right). \label{Red150}
\end{equation}

We are now equipped to compute the corresponding radiated energy in the case 
under consideration. 

In order to accomplish this purpose, let us start by observing that the density of power carried out by the radiation fields across the surface bounding the volume $V$ is given by the real part of the Poynting vector (time averaged value)
\begin{equation}
{\bf S} =\frac{c}{{2\pi }}{\mathop{\rm Re}\nolimits} \left( {{\bf e} \times {\bf {h^ * }}} \right). \label{Red155}
\end{equation}

We further recall that we will calculate the power radiated through the surface \cite{Das}, that is,
 \begin{equation}
{\cal E} = \int_{ - \infty }^\infty  {dt} \int\limits_S {d{\bf a} \cdot {\bf S}}.  \label{Red160}
\end{equation}
It is worth emphasizing that in our case we shall consider a cylinder as the integration surface. Also, it may be mentioned that in order to get a meaningful expression we shall use a cylinder infinitesimally small \cite{PRA}. 

Consequently, the power radiated per unit length through the surface
 then reads
\begin{eqnarray}
{\cal E} &=& \frac{c}{{2\pi }}{\mathop{\rm Re}\nolimits} \int_0^\infty  {dw} \left\{ {2\pi {\rho _0}{{\left. {{S_\rho }} \right|}_{\rho  = {\rho _0}}}} \right\}  \nonumber\\
&+& \frac{c}{{2\pi }}{\mathop{\rm Re}\nolimits} \int_0^\infty  {dw} \left\{ {\frac{\partial }{{{\partial _z}}}\int_0^{{\rho _0}} {\int_0^{2\pi } {{S_z}} \,\rho\, d\rho\, d\phi } } \right\}, \label{Red165}
\end{eqnarray}
where ${S_\rho } =  - {e_z}h_\phi ^ *$, ${S_z} = {e_\rho }h_\phi ^ *$ and ${\rho _0} \to 0$. Let us also recall here that, in our case, the $\phi$-component of the Poynting vector $\left( {{S_\phi }} \right)$ vanishes.

According to equations (\ref{Red125}), (\ref{Red145}) and (\ref{Red150}), the expression for the power radiated per unit length (\ref{Red165}) takes the form
\begin{equation}
{\cal E} = - \pi\, \frac{{{Q^2}v}}{{{c^2}}}\,\frac{{{n^2}}}{\varepsilon }\int_0^\infty  {dww\left( {1 - \frac{{{c^2}}}{{{n^2}{v^2}}}} \right)}. \label{Red170}
\end{equation}
One immediately sees that this expression is similar to that encountered in the Cherenkov radiation theory \cite{Das}. This last expression clearly shows the role played by vacuum electromagnetic nonlinearities in triggering the radiated energy. We also point out that in equation (\ref{Red170}) we have used the asymptotic behavior of the Bessel (${K_\nu }\left( x \right) \to \frac{\pi }{{\sqrt {2x} }}{e^{ - x}}$), since we are describing outgoing radiation.

In connection with this last expression (\ref{Red170}) a few comments are in order. First, it should be recalled that ${\cal E}$ represents the rate of energy lost due to radiation along the trajectory of the charged particle, $-\frac{{dE}}{{dt}}$, where $E$ is the energy of the charged particle. Second, we also recall that the $w$ integration has physical meaning only over the range where $n > {c \mathord{\left/
 {\vphantom {c v}} \right.\kern-\nulldelimiterspace} v}$. Third, we further note that the vacuum of the electrodynamics studied in this work describes a non-dispersive ''medium'', which is verified because $\varepsilon$ and $\mu$ are constant. In other words, the velocity of electromagnetic waves in this ''medium'' does not depend on the frequency of the waves.\\
 
\section{Final remarks}

In summary, we have studied some observational signatures of nonlinearities of the electromagnetic field. First, we shown the vital role played by nonlinearities in triggering a material with  $(\varepsilon > 0, \mu <0)$, which corresponds to a ferrimagnetic material. Secondly, the permittivity and susceptibility induced by nonlinearities have been studied in order to obtain the refractive index via the dispersion relation for logarithmic electrodynamics. Finally, we have considered 
the radiation produced by a moving charged particle (with uniform velocity) interacting in nonlinear medium. Let us also recall here that particles moving with a uniform velocity in vacuum do not lead to radiation. As already mentioned, one is lead to the interesting conclusion that the above radiation is driven by the medium through which the particle travels like the one that happens in the Cherenkov effect. Lastly, we will be focusing efforts to understand in more detail the physical consequences of electrodynamics with metamaterial features, including Cherenkov radiation, in the near future.

\section*{Acknowledgments}

One of us (P. G.) was partially supported by ANID PIA / APOYO AFB180002.

\end{document}